\documentclass[conference]{IEEEtran}
\IEEEoverridecommandlockouts
\usepackage{cite}
\usepackage{amsmath,amssymb,amsfonts}
\usepackage{algorithmic}
\usepackage{graphicx}
\usepackage{booktabs}
\usepackage{makecell}
\usepackage{multirow}
\usepackage{textcomp}
\usepackage{xcolor}
\usepackage[colorlinks  = true,
            linkcolor   = black,
            urlcolor    = magenta,
            anchorcolor = blue,
            bookmarks   = false
            ]{hyperref}

\def\BibTeX{{\rm B\kern-.05em{\sc i\kern-.025em b}\kern-.08em
    T\kern-.1667em\lower.7ex\hbox{E}\kern-.125emX}}
\graphicspath{ {./images/} }

\DeclareMathOperator*{\LeakyReLU}{LeakyReLU}

\begin{document}

\title{Multi-resolution CSI Feedback with deep learning in Massive MIMO System \\
}

\author{\IEEEauthorblockN{Zhilin Lu}
\IEEEauthorblockA{\textit{Beijing National Research Center for}\\ \textit{Information Science and Technology}\\ \textit{(BNRist), Tsinghua University}\\
Beijing, China \\
luzl18@mails.tsinghua.edu.cn}\and

\IEEEauthorblockN{Jintao Wang}
\IEEEauthorblockA{\textit{Beijing National Research Center for}\\ \textit{Information Science and Technology}\\ \textit{(BNRist), Tsinghua University}\\
Beijing, China \\
wangjintao@tsinghua.edu.cn}\and

\IEEEauthorblockN{Jian Song}
\IEEEauthorblockA{\textit{Beijing National Research Center for}\\ \textit{Information Science and Technology}\\ \textit{(BNRist), Tsinghua University}\\
Beijing, China \\
jsong@tsinghua.edu.cn}\and
}
\maketitle

\begin{abstract}
In massive multiple-input multiple-output (MIMO) system, user equipment (UE) needs to send downlink channel state information (CSI) back to base station (BS). However, the feedback becomes expensive with the growing complexity of CSI in massive MIMO system. Recently, deep learning (DL) approaches are used to improve the reconstruction efficiency of CSI feedback. In this paper, a novel feedback network named CRNet is proposed to achieve better performance via extracting CSI features on multiple resolutions. An advanced training scheme that further boosts the network performance is also introduced. Simulation results show that the proposed CRNet outperforms the state-of-the-art CsiNet under the same computational complexity  without any extra information. The open source codes are available at  \textnormal{\href{https://github.com/Kylin9511/CRNet}{https://github.com/Kylin9511/CRNet}}
\end{abstract}

\begin{IEEEkeywords}
Massive MIMO, CSI feedback, deep learning, convolutional neural network, inception network
\end{IEEEkeywords}

\section{Introduction}
Massive multiple-input multiple-output (MIMO) is a promising technique to improve the spectrum and energy efficiency for the next generation wireless systems \cite{01massivemimo, 02massivemimo}. However, new challenge appears that base station (BS) needs to obtain the real time channel state information (CSI) for precoding. The uplink CSI can be acquired by channel estimation, while the downlink CSI has to be fed back from user equipment (UE) especially in frequency division duplexing (FDD) systems \cite{03massivemimo}. The feedback is challenging in massive MIMO system due to the huge CSI matrix. The bandwidth overhead of CSI feedback becomes unacceptable.

In order to reduce the feedback overhead, the CSI matrix must be compressed. The traditional methods based on compressed sensing (CS)\cite{05cs} require the CSI matrix to be sparse enough. However, the practical system can not meet the requirement especially when the compression ratio is large.

On the other hand, artificial intelligence (AI) and deep learning (DL) have caught wide attention in the past few years. Many neural network (NN) backbones including ResNet \cite{06resnet} and InceptionNets\cite{07inceptionv1, 09inceptionv3, 10inceptionv4} are proposed. They are proved to be robust and effective in many tasks including image compression\cite{11reconnet, 12dr2net, 13imagerecovery}, which is similar to CSI compression. This inspires the researchers to solve the CSI feedback problem via DL based encoder-decoder network.

Recently, a series of works have been introduced on downlink Massive MIMO CSI feedback with DL. A neural network named ``CsiNet'' is designed in \cite{14csinet} and its overwhelming superiority against traditional CS methods is given. After that, some researchers expand the original scene. Correlation of CSI in different time slots is used in \cite{15csinet_time_varying} and correlation between the uplink and downlink CSI is utilized in \cite{16csinet_mapping, 17csinet_mapping2}. These can be seen as new sub-scenes with extra conditions or assumptions.

The CsiNetPlus proposed in \cite{18csinet_plus} improves the network performance without extra information by updating the convolutional kernels. It proves the huge potential for network design in pure CSI feedback task. However, CsiNetPlus inherits most of the CsiNet architecture design. What is more, the floating point operations (flops) of CsiNetPlus is much larger, making the improvement of CsiNetPlus sort of a complexity-performance trade off.

In this paper, we propose a novel neural network called channel reconstruction network (CRNet) based on multi-resolution architecture. We also introduce an advanced training scheme that fit the CSI feedback task better to boost the performance. Simulation results show that with the same computational complexity, the proposed CRNet greatly outperforms the CsiNet especially under the advanced scheme.

The main contribution of this paper is listed below.

\begin{itemize}
	\item Multi-resolution CRBlock is designed based on inception block \cite{07inceptionv1}. With the help of CRBlock, the CRNet is able to extract multi-resolution features and adapt to various scenario and compression ratio. 
	\item We adopt the convolution factorization in \cite{09inceptionv3} and prove its effectiveness in CSI feedback task. It can expand the resolution without any flops increasing.
	\item We introduce warm up aided cosine learning rate (lr) scheduler. The effectiveness of exhaustive training is also analyzed. To the author's best knowledge, we are the first to emphasize the importance of training scheme on CSI feedback task. Advanced scheme is proposed and proved to be practical.
\end{itemize}

The rest of the paper is organized as follows. Section \ref{Section-SystemModel} introduces the system model and the CSI feedback scenario. Section \ref{Section-Design} explains the detailed design of CRNet and the advanced training scheme. After that the numerical results and analysis are presented in Section \ref{Section-Results}. The final conclusion is drawn in Section \ref{Section-Conclusion}.

\section{System Model} \label{Section-SystemModel}
In this work, we consider a single cell massive MIMO FDD system with $N_t$ antennas at BS and $N_r$ antennas at UE. In this scenario we have $N_t \gg 1$, and $N_r$ is set to $1$ for simplicity. Orthogonal frequency division multiplexing (OFDM) with $N_c$ sub-carrier is adopted, where the received signal $\mathbf{y} \in \mathbb{C}^{N_c \times 1}$ can be derived as follows:

\begin{equation}
	\mathbf{y} = \mathbf{A}\mathbf{x} + \mathbf{z},
\end{equation}
where $\mathbf{x} \in \mathbb{C}^{N_c \times 1}$ is the transmitted symbol vector in one OFDM period, and $\mathbf{z} \in \mathbb{C}^{N_c \times 1}$ is the additive noise vector. Diagonal channel matrix $\mathbf{A} = \text{diag}(\mathbf{h}_1^H\mathbf{p}_1, \cdots, \mathbf{h}_n^H\mathbf{p}_n)$ where $n=N_c$, $\mathbf{h}_i \in \mathbb{C}^{N_t \times 1}$ and $\mathbf{p}_i \in \mathbb{C}^{N_t \times 1}, i \in \{1, \cdots, N_c\}$ are downlink channel response vector and beamforming vector for each sub-carrier, respectively.

In order to design the beamforming vector $\mathbf{p}_i$, the BS needs to acquire the corresponding $\mathbf{h}_i$. We can define the downlink channel matrix as $\mathbf{H}=[\mathbf{h}_1\cdots \mathbf{h}_{N_c}]^H$ which contains $N_cN_t$ elements. Typically  $N_cN_t$ for massive MIMO FDD system is unacceptably large.

Luckily the channel matrix $\mathbf{H}$ is sparse in the angular-delay domain \cite{14csinet}. Following equation (\ref{eq2}), we can transfer the channel matrix from spatial-frequency domain to angular-delay domain via discrete Fourier transform (DFT).

\begin{equation} \label{eq2}
	\mathbf{H}' = \mathbf{F}_c\mathbf{H}\mathbf{F}_t^H,
\end{equation}
where $\mathbf{F}_c$ and $\mathbf{F}_t$ are the DFT matrices with dimension $N_c\times N_c$ and $N_t\times N_t$, respectively. For angular-delay domain channel matrix $\mathbf{H}'$, only the first $N_a$ row contains large values. The rest of the rows are made up of near zero elements that can be left out without much information loss. For easy understanding, we use $\mathbf{H}_a$ to denote the first $N_a$ rows of $\mathbf{H}'$.

Although $\mathbf{H}_a$ is much lighter compared with $\mathbf{H}$, $N_aN_t$ is still a large number. That is why we need to further compress $\mathbf{H}_a$ before  feedback. $\mathbf{H}_a$ is sparse enough for the CS based methods when $N_t \rightarrow \infty$. However, $N_t$ is limited in the practical system, making the sparsity of $\mathbf{H}_a$ insufficient especially for large compression ratio. The NN based approaches can overcome the shortcoming of CS and achieve better results.

\begin{figure}[t]
\centering
\includegraphics[width=0.5\textwidth]{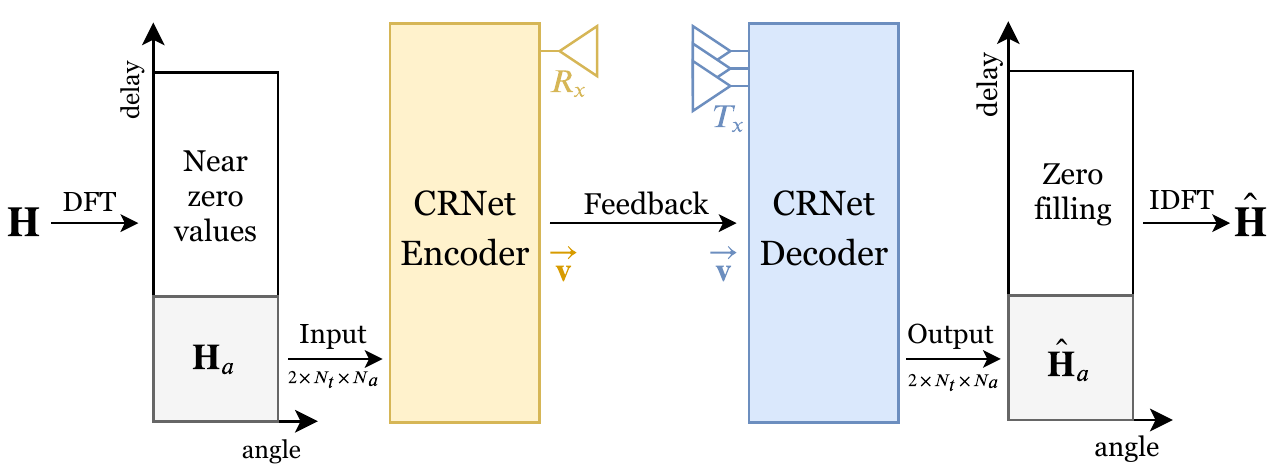}
\caption{Schematic diagram of CRNet aided downlink CSI feedback workflow}
\label{System Model}
\end{figure}

In this paper we consider an encoder-decoder network for the downlink SCI feedback. As it is shown in Fig. \ref{System Model}, the channel matrix $\mathbf{H}$ is first transferred to angular domain by DFT. Then the encoder of CRNet compresses $\mathbf{H}_a$ into a short feature vector $\mathbf{v}$ according to a given compression ratio $\eta$. After $\mathbf{v}$ is fed back to BS, it will be reconstructed into $\mathbf{H}_a$ by the decoder of CRNet. Finally $\mathbf{H}$ can be restored by zero filling and inverse DFT.

The whole feedback scheme can be concluded by (\ref{eq3}).

\begin{equation} \label{eq3}
	\hat{\mathbf{H}}_a = \mathcal{D}(\mathcal{E}(\mathbf{H}_a, \Theta_\mathcal{E}), \Theta_\mathcal{D}),
\end{equation}
where $\mathcal{E}$ and $\mathcal{D}$ denote the encoder and the decoder of CRNet, respectively. $\Theta_\mathcal{E}$ and $\Theta_\mathcal{D}$ represent their network parameters. Our purpose is to design and train $\Theta_\mathcal{E}$ and $\Theta_\mathcal{D}$ so that the distance between $\mathbf{H}_a$ and the reconstructed $\hat{\mathbf{H}}_a$ is minimized.

Note that this work only focuses on the feedback scheme, the downlink channel estimation and the uplink feedback are assumed to be ideal. Besides, we adopt COST2100 \cite{19Cost2100} model to simulate the channel matrix $\mathbf{H}$ for the massive MIMO FDD system.

\section{Design of CRNet and Training Scheme} \label{Section-Design}
\subsection{The proposed CRNet} \label{SubSection-CRNet}

Pure residual architecture, like RefineNet in CsiNet, is applied and proved to be effective in previous works. It extracts residual CSI features under a fixed resolution. However, the sparsity of CSI varies according to the channel scenario and the compression ratio in our task. 

If the CSI matrix is relatively dense, its feature granularity will be finer. Convolution with smaller kernel size can extract finer features better. Conversely, larger kernel is preferred for sparser CSI matrix. Single resolution network can not adapt to the feature granularity change well, which may leads to loss in network performance and robustness.

\begin{figure}[t]
\centering
\includegraphics[width=0.49\textwidth]{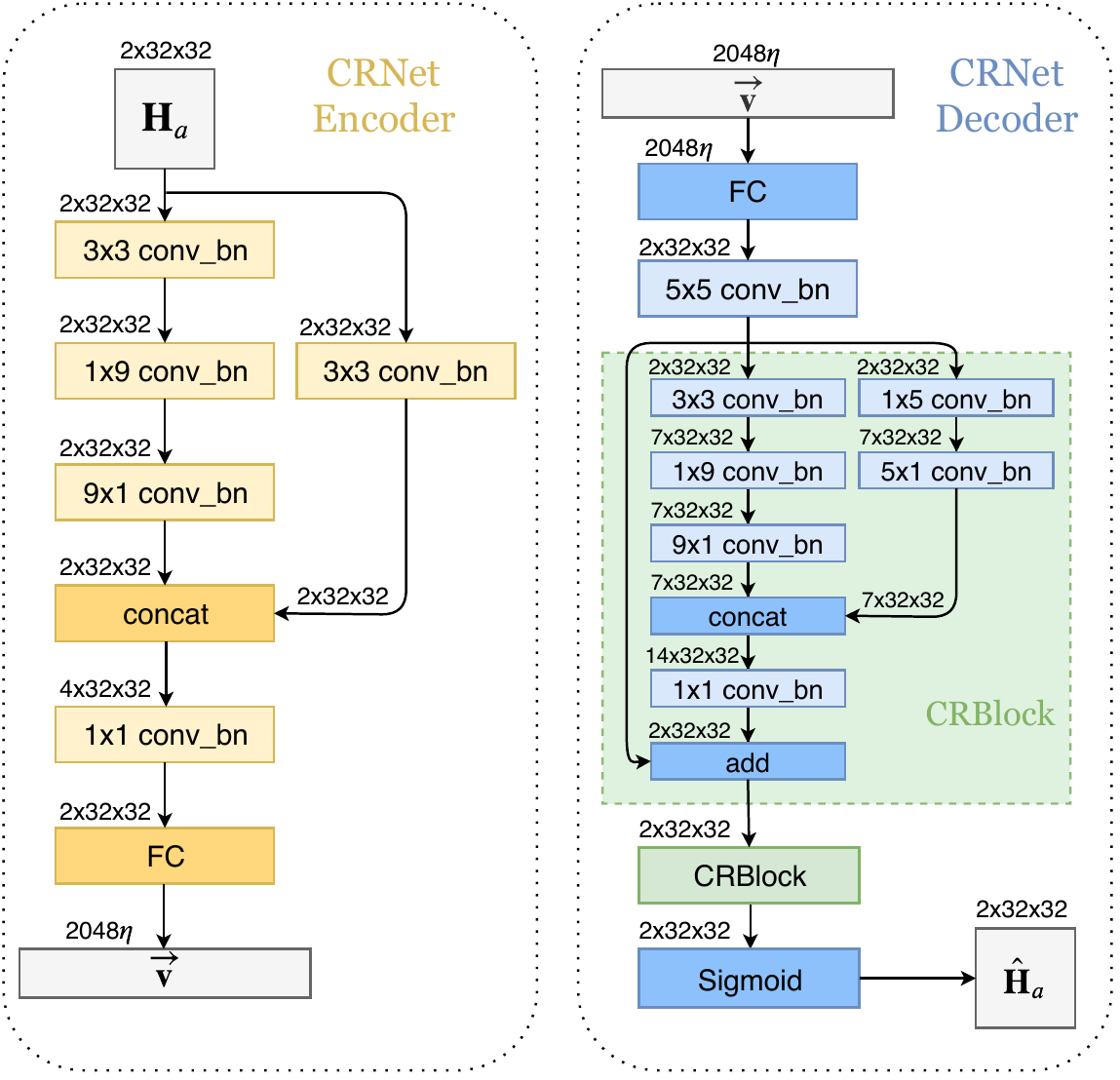}
\caption{Encoder and decoder design of the proposed CRNet. All the input feature shape ($c\times h \times w$) is given on top of the corresponding block. The activation functions and reshape blocks are left out in the diagram for simplicity.}
\label{CRNet}
\end{figure}

Considering the aforementioned observation, we introduce multi-resolution network to the CSI feedback task. A brand new network named CRNet is proposed and its architecture is demonstrated in Fig. \ref{CRNet}. CRNet consists of two separated parts: the encoder at UE and the decoder at BS. The angular channel matrix $\mathbf{H}_a$ is treated as an input image of size $2\times N_a \times N_t$, and its two channels correspond to the real and imaginary part of $\mathbf{H}_a$.

For the encoder, input image will pass through two parallel paths. One path is made up of three serial convolution layers, providing large resolution view. The other path only contains one $3\times 3$ convolution layer whose resolution is much smaller. Then we concatenate the outputs and merge them with a $1\times 1$ convolution. Finally a fully connected (FC) layer is used to scale down the feature based on the given compression ratio.

For the decoder, the received feature vector $\mathbf{v}$ is first scaled up and resized. After that, a convolution layer extract the feature roughly. Then it comes to two CRBlocks, which is the core design of the decoder. Each CRBlock consists of two parallel paths with different resolutions. Then a $1\times 1$ convolution layer merges two output features. Identity path is added to each CRBlock following the idea of residual learning. At the end, there is an extra sigmoid layer for value range adjustment and further activation.

Note that each convolution is followed by a batch normalization as it is depicted in Fig. \ref{CRNet}. Besides, a leaky ReLU activation layer is added at the end of each ``conv\_bn'' layer to provide nonlinearity. Also, reshape operation is added where it is necessary. The definition of leaky ReLU is shown in (\ref{eq4}).

\begin{equation} \label{eq4}
	\LeakyReLU(x) = \left\{
		\begin{array}{ll}
			x, & x \ge 0 \\
			\beta x, & x < 0,
		\end{array} 
	\right.
\end{equation}
where $\beta \in (0, 1)$ is the negative slope. We set $\beta$ to 0.3 in the proposed CRNet.

It is obvious that the multi-resolution CRBlock is the key to the CRNet design. In fact, the encoder is a simplified CRBlock as well. By setting two different convolutional paths in parallel, the proposed CRBlock can flexibly extract the features from different sizes. In this way the network capacity of adapting to different feature granularities are enhanced.

What is more, convolution factorization is first introduced to CSI feedback task and proved to be effective. By replacing a huge $9\times 9$ kernel with a $1\times 9$ kernel and a $9\times 1$ kernel in series, the resolution area is kept while the computational complexity is largely reduced.

With the help of multi-resolution paths and convolution factorization, CRNet achieves better performance under the same or even less computational complexity compared with CsiNet.

\subsection{The advanced training scheme design} \label{SubSection-Scheme}

The importance of training scheme design is always underestimated in previous DL aided CSI feedback works. CsiNet is trained for 1000 epochs with a fixed learning rate of 0.001\cite{14csinet}. But we can not find further ablation study to explain why that scheme is good. Some other works even omit the basic training scheme description in their essay, which seriously harm the reliability of their reported network performance. 

In fact, no network can produce good result if it is not trained properly. Besides, the training scheme has no impact on online inference cost, making the benefit it brings free of charge. Generally speaking, the best training scheme is different for different tasks and networks. Therefore it is important to explore the best scheme for brand new task like CSI feedback. 

In this paper, we mainly look into two aspects of training scheme design: the number of training epochs and the learning rate (lr) scheduler.

In order to set the training epoch rationally, we should pay attention to the overfitting problem. Generally speaking, if the task is unlikely to overfit, longer training will help finding a better solution. However, for the majority of tasks in computer vision, overfitting happens easily when training is long and it harms the network performance. Therefore we need to analyze the possibility of overfitting in CSI feedback task.

As a matter of fact, overfitting is quite hard for the CSI feedback task. There are mainly two reasons for it. For one thing, the model we use is fairly simple compared with models in traditional computer vision. For example, flops of the famous ResNet50 is 3.9G, which is over 500 times larger than our proposed CRNet. For the other thing, the data we use does not contain complicated yet irrelevant information. An image of a dog may contains its owner's hand, the house, the grass, etc. However, the property of CSI matrix is very special. It is random but descriptive, even the noise can be formatted with reasonable assumption. 

Overfitting is more likely to happen when the network is too powerful and the data contains intertwined information. Apparently CRNet is very light and the information in the CSI matrix is rather pure. That means the main challenge would be under fitting especially for large compression ratio. Based on the aforementioned analysis, we suggest exhausted training for CSI feedback network. Experiments shows that training long enough is helpful indeed.

\begin{figure}[t]
\centering
\includegraphics[width=0.5\textwidth]{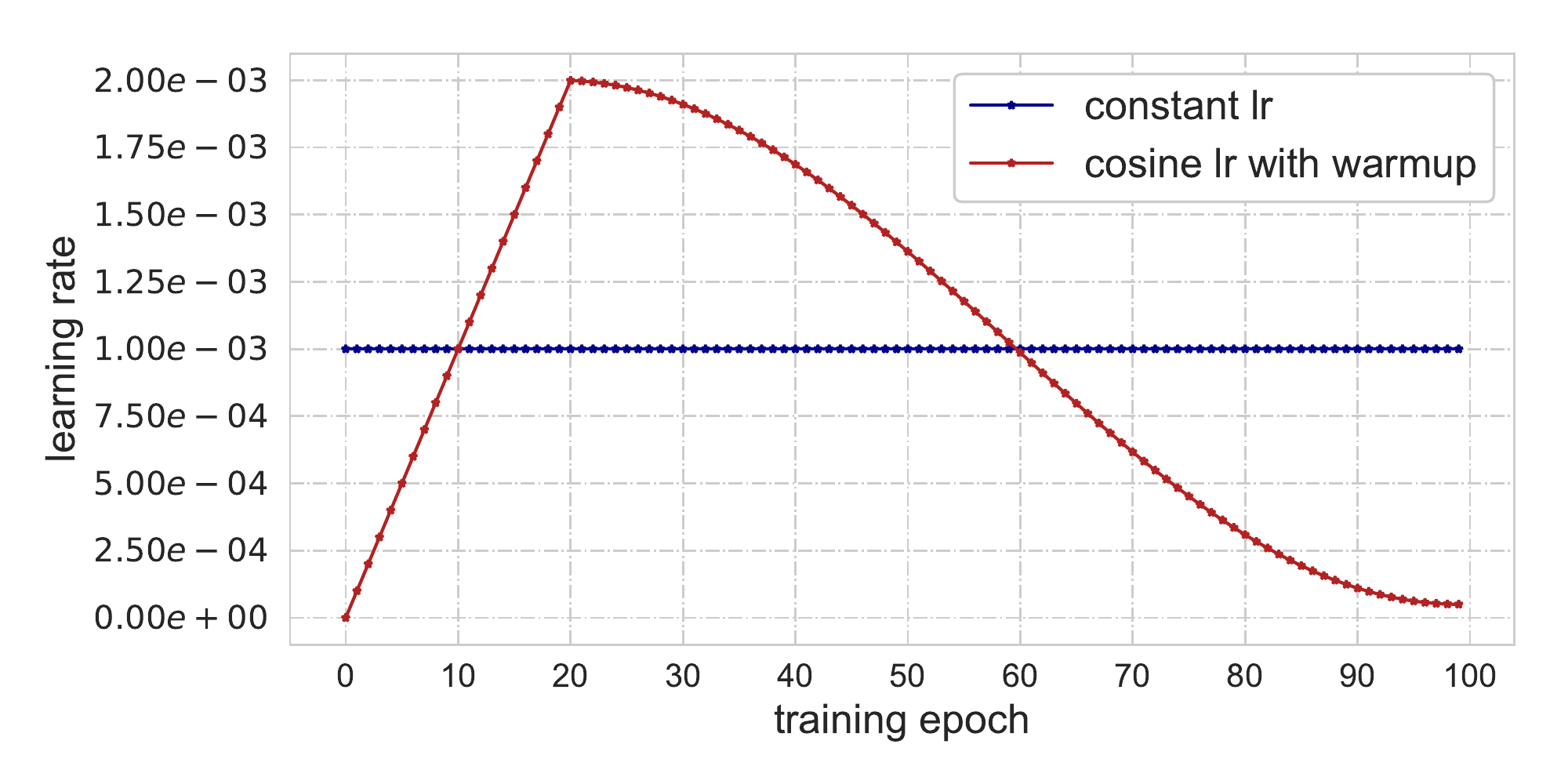}
\caption{Comparison between constant lr scheduler and cosine annealing lr scheduler with linear warm up.}
\label{scheduler}
\end{figure}

The scheduler design is another key problem. As it is well known, a proper lr scheduler is important for stochastic gradient descent (SGD) optimizer. For instance, cosine annealing lr with restart is proved to be significantly better compared with constant lr when using SGD opimizer \cite{20sgdr}.

Adam optimizer is less sensitive to lr scheduler since it can auto-adapt lr during the optimizing process. However, our experiment shows that for CSI feedback task the constant lr is not enough even for Adam. By applying cosine annealing lr with warm up depicted in Fig. \ref{scheduler}, we significantly improve the performance of CRNet. 

To be specific, the lr is linearly increased from zero to the maximal rate at the beginning of training. This operation is called "warm up" and the network parameters are updated to a better initial position. After that the lr decays from maximal rate following (\ref{eq5}).

\begin{equation} \label{eq5}
	\gamma = \gamma_{min} + \frac{1}{2}\left(\gamma_{max} - \gamma_{min}\right) \left(1 + \cos\left(\frac{t - T_w}{T - T_w}\pi\right)\right)_,
\end{equation}
where $\gamma$, $\gamma_{max}$ and $\gamma_{min}$ are current, initial and final lr, respectively. $t$ is index of current epoch, $T_w$ and $T$ are the number of warm up and total epochs, respectively.

Here comes the reason why cosine annealing lr works better than constant lr. At the early stage, network needs to be trained long enough with large lr to locate a better zone. When it comes to the final stage, training longer with small lr usually means getting closer to optimum value. As we can see in Fig. \ref{scheduler}, the initial lr is 40 times large than the final lr for cosine scheduler. The early stage and final stage are relatively longer than the middle stage due to the shape of cosine function. Besides, lr update is smooth and continuous under cosine scheduler, making training more stable.

In conclusion, we propose an advanced training scheme to train CRNet in a better way. With the help of warm up aided cosine annealing lr scheduler and exhausted training, our advanced scheme significantly improves the network performance.

\section{Simulation Results and Analysis} \label{Section-Results}
\subsection{Experiment Setting} \label{Subsection-experiment-setting}

We consider two types of scenarios: the indoor scenario at 5.3GHz and the outdoor scenario at 300MHz. The channel is generated following the default setting in COST2100 \cite{19Cost2100}. We adopt the basic system settings in CsiNet \cite{14csinet} for easier comparison. Uniform linear array (ULA) model with $N_t=32$ is considered at BS. For FDD system, we take $N_c=1024$ in the frequency domain and $N_a=32$ in the angular domain. 150,000 independently generated channels are split into three parts. The training, validation and test dataset contain 100,000, 30,000 and 20,000 channel matrices, respectively. The batch size is set to 200.

The whole pipeline is implemented in PyTorch. The Xavier initialization is applied on both convolution layers and FC layers. We use Adam optimizer with default setting ($\beta_1=0.9, \beta_2=0.999, \epsilon=1e-8$) and mean square error (MSE) loss. For our advanced training scheme, the initial and final lr of cosine annealing scheduler is set to be 2e-3 and 5e-5. The network is trained for 2500 epochs which is long enough according to our observation. The first 30 epochs is used for warm up.

\subsection{Performance of the proposed CRNet and training scheme}

\begin{table}[!b]
\caption{NMSE (dB) and flops comparison between CsiNet and CRNet}
\begin{center}
\begin{tabular}{c c|c c|c c}
\Xhline{0.8pt}
\multirow{2}{*}{$\mathbf{\eta}$} & \multirow{2}{*}{\textbf{Methods}} & \multicolumn{2}{c|}{\textbf{Indoor}} & \multicolumn{2}{c}{\textbf{Outdoor}}\\
	& & NMSE & flops & NMSE & flops\\
\Xhline{0.8pt}
\multirow{3}{*}{1/4} & CsiNet & -17.36 & 5.41M & -8.75 & 5.41M \\
	& CRNet-const & -21.17 & 5.12M & -10.42 & 5.12M \\
	& CRNet-cosine & \textbf{-26.99} & 5.12M & \textbf{-12.71} & 5.12M \\
\hline
\multirow{3}{*}{1/8} & CsiNet & -12.70 & 4.37M & -7.61 & 4.37M \\
	& CRNet-const & -13.79 & 4.07M & -7.67 & 4.07M \\
	& CRNet-cosine & \textbf{-16.01} & 4.07M & \textbf{-8.04} & 4.07M \\
\hline
\multirow{3}{*}{1/16} & CsiNet & -8.65 & 3.84M & -4.51 & 3.84M \\
	& CRNet-const & -10.29 & 3.55M & -5.09 & 3.55M \\
	& CRNet-cosine &  \textbf{-11.35} & 3.55M & \textbf{-5.44} & 3.55M \\
\hline
\multirow{3}{*}{1/32} & CsiNet & -6.24 & 3.58M & -2.81 & 3.58M \\
	& CRNet-const & -8.58 & 3.28M & -3.19 & 3.28M \\
	& CRNet-cosine & \textbf{-8.93} & 3.28M & \textbf{-3.51} & 3.28M \\
\hline
\multirow{3}{*}{1/64} & CsiNet & -5.84 & 3.45M & -1.93 & 3.45M \\
	& CRNet-const & -6.14 & 3.16M & -2.13 & 3.16M \\
	& CRNet-cosine & \textbf{-6.49} & 3.16M & \textbf{-2.22} & 3.16M \\
\Xhline{0.8pt}
\end{tabular}
\label{tab1}
\end{center}
\end{table}

\begin{table}[!b]
\caption{Impact of training epochs on CRNet NMSE (dB) performance}
\begin{center}
\begin{tabular}{c|c c c c c}
\Xhline{0.8pt}
\multirow{2}{*}{\textbf{\shortstack{Training \\ scheme$^{\mathrm{a}}$}}} & \multicolumn{5}{c}{\textbf{Number of training epochs}} \\
 & 100 & 500 & 1000 & 2500 & 5000 \\

\Xhline{0.8pt}

CRNet-const  & -18.94 & -17.55 & -21.17  & -22.94 & -21.89 \\
CRNet-cosine & -21.33 & -23.76 & -25.17  & -26.99 & \textbf{-27.31} \\

\Xhline{0.8pt}
\multicolumn{6}{l}{$^{\mathrm{a}}$All schemes are tested under the indoor scenario ($\eta=1/4$).}
\end{tabular}
\label{tab2}
\end{center}
\end{table}

In order to evaluate the performance, we measure the distance between the original $\mathbf{H}_a$ and the reconstructed $\hat{\mathbf{H}}_a$ with normalized mean square error (NMSE) defined in (\ref{eq6}).

\begin{equation} \label{eq6}
	\text{NMSE} = E\left\{\frac{\Vert \mathbf{H}_a - \hat{\mathbf{H}}_a \Vert_2^2} {\left\Vert \mathbf{H}_a\right\Vert_2^2} \right\}
\end{equation}

The main performance comparison between the original CsiNet and our proposed CRNet is listed in Table \ref{tab1}. The ``CRNet-const'' stands for CRNet trained with the same scheme as CsiNet, which is 1000 epochs of training under constant learning rate 0.001. And the ``CRNet-cosine'' represent CRNet trained with our advanced scheme described in Section \ref{Subsection-experiment-setting}.

Table \ref{tab1} shows that under the same training scheme, CRNet stably outperform CsiNet with less flops, demonstrating the advantage of CRNet architecture design. The multi-resolution CRBlock with convolution factorization is proved to be suitable for the CSI feedback t ask. What is more, the performance of CRNet further improves under our advanced training scheme without any extra flops. The warm up and cosine annealing scheduler introduced in Section \ref{SubSection-Scheme} do work for the scene.

\begin{figure}[!t]
\centering
\includegraphics[width=0.5\textwidth]{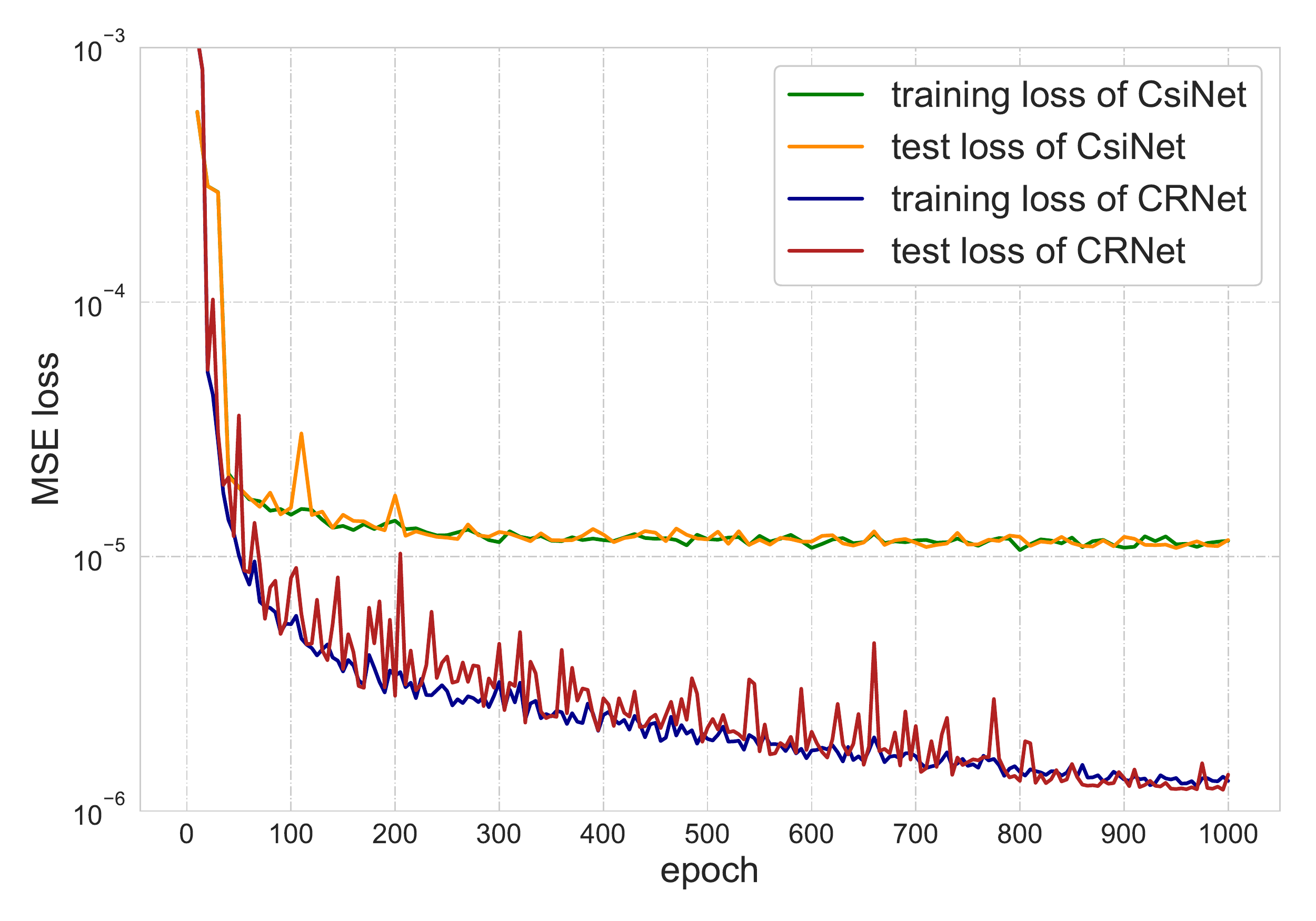}
\caption{Loss descending trends comparison between original CsiNet and the proposed CRNet. Note that the CRNet is trained under cosine annealing scheduler with warm up for 1000 epochs.}
\label{loss}
\end{figure}

Another important problem is the best training epoch setting. Ablation study in Table \ref{tab2} shows that the performance generally gets better with the increase of training epochs, which matches with our analysis in Section \ref{SubSection-Scheme}.

There is an exception that NMSE of CRNet-const seems to be worse when it is trained for 5000 epochs. However, that is due to the instability of training under constant scheduler. Sometimes, the training happens to be trapped into a zone with relatively bad local minimum, which is why we need a better training scheme. The exhausted training is effective when you consider one single experiment. For instance, when the unluckily trapped network is trained for 2500 epochs, its NMSE is merely -19.18 dB. By training for 2500 more epochs, its NMSE decrease for another 2.71 dB even in a bad local zone.

Exhausted training only works for the tasks that are unlikely to overfit. Since the CSI feedback task can benefit from exhausted training, the overfitting of it must be slight. We can establish an intuitive concept from Fig. \ref{loss}, where loss descending trends of training and test are presented. CRNet is trained with warm up aided cosine scheduler for only 1000 epochs in Fig. \ref{loss} for easier comparison. For both CsiNet and CRNet, the test loss fluctuates around the training loss all the way, suggesting a relatively weak overfitting during training. 

However, the benefit of longer training decreases as the training gets longer. According to our experiments, CRNet can perform well enough under cosine scheduler with 2500 epochs of training. That is why we set the number of training epoch to 2500 in our final advanced scheme. 

\begin{figure}[!t]
\centering
\includegraphics[width=0.5\textwidth]{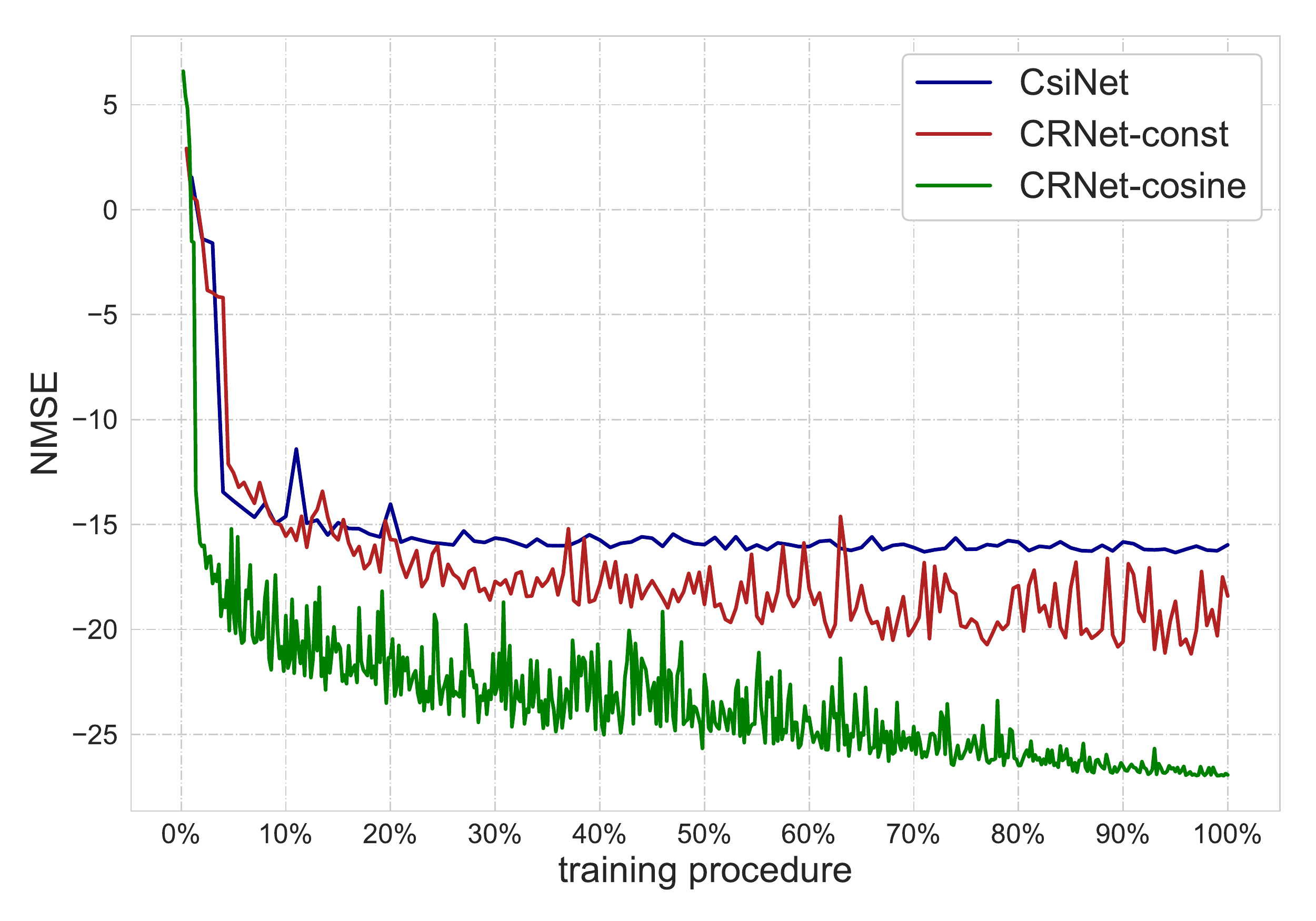}
\caption{NMSE trends of CsiNet and CRNet under different schemes. Note that the horizontal axis is set to training procedure for clearer comparison. CsiNet and CRNet-const are both trained for 1000 epochs with learning rate 0.001. CRNet-cosine is trained for 2500 epochs with cosine annealing lr, which is our advanced scheme.}
\label{nmse}
\end{figure}

The specific difference of the loss curves between the advanced scheme aided CRNet and the original CsiNet can also be revealed by Fig. \ref{loss}. The loss of CRNet decreases sharply for a longer time. What is more, it is still decaying slowly when the loss of CsiNet is stuck on plateau. Besides, a clear comparison of NMSE training curves can be found in Fig. \ref{nmse}. As we can see, under the same training scheme, our proposed CRNet reaches lower NMSE with relatively higher fluctuation. And with the help of cosine annealing scheduler and exhausted training, The NMSE of CRNet is significantly lower. Besides, the NMSE curve gets less fluctuant as the training goes, suggesting that the result is more stable under the new scheme.

\begin{table}[b]
\caption{NMSE(dB) performance for different negative slope on CRNet}
\begin{center}
\begin{tabular}{c|c c c c c}
\Xhline{0.8pt}
\multirow{2}{*}{\textbf{\shortstack{Scenarios$^{\mathrm{a}}$}}} & \multicolumn{5}{c}{\textbf{Leaky ReLU negative slope}} \\
 & 0.01 & 0.05 & 0.1 & 0.3 & 0.5 \\

\Xhline{0.8pt}

Indoor   & -14.79 & -23.06 & -23.45  & \textbf{-26.99}  & -25.39 \\
Outdoor  & -8.99  & -10.74 & -10.63  & \textbf{-12.71}  & -12.54\\

\Xhline{0.8pt}
\multicolumn{6}{l}{$^{\mathrm{a}}$The compression ratio $\eta$ is set to $1/4$ for all scenarios.}
\end{tabular}
\label{tab3}
\end{center}
\end{table}

\begin{table}[b]
\caption{Performance of different head convolution in CRNet decoder}
\begin{center}
\begin{tabular}{c|c c|c c}
\Xhline{0.8pt}
\multirow{2}{*}{\textbf{\shortstack{Head\\ convolution}}} & \multicolumn{2}{c|}{\textbf{Indoor ($\eta=1/4$)}} & \multicolumn{2}{c}{\textbf{Outdoor ($\eta=1/4$)}}\\
 & NMSE & flops & NMSE & flops\\

\Xhline{0.8pt}

Blank    & -25.55 & 5.02M & -12.5 & 5.02M \\
$3\times 3$      & -24.9 & 5.06M & -11.3 & 5.06M \\
Dual $3\times 3$ & -25.9 & 5.09M & -11.9 & 5.09M \\
$5\times 5$      & \textbf{-26.99} & 5.12M & \textbf{-12.71} & 5.12M \\
$7\times 7$      & -26.31 & 5.22M & -11.84 & 5.22M \\

\Xhline{0.8pt}
\end{tabular}
\label{tab4}
\end{center}
\end{table}

Finally, we will offer some interesting observations. We test the CRNet performance with different leaky ReLU negative slopes and the results are shown in Table \ref{tab3}. When leaky ReLU is used in some computer vision tasks, small negative slope can also work well. However, larger slope is essential to boost the performance of CRNet. The optimization of CRNet will be much harder if the negative slope is too small.

Besides, we investigate into the head convolution design in CRNet and present the results in Table \ref{tab4}. Generally speaking, neural network based on pure residual architecture needs a head convolution layer to roughly extract the feature. For example, at the beginning of ResNet \cite{06resnet} there exists a $7\times 7$ convolution layer. In CsiNetPlus \cite{18csinet_plus}, adding head convolution layer to the original CsiNet is listed as one of the core upgrades. 

However in CRNet, we find that the head convolution layer has little influence on the final performance. For indoor scenario with $\eta=4$, NMSE only decreases for around 1.5dB if you remove the $5\times 5$ head convolution layer. Besides, the performance of different head layers are relatively similar. This may thanks to the multi-resolution architecture in CRNet, which can better adapt to the change of CSI features. Still we choose to add a $5\times 5$ head convolution layer into CRNet for better performance, since the flops it consumes is negligible.

\section{Conclusion} \label{Section-Conclusion}

In this paper, a novel neural network named CRNet was proposed for downlink CSI feedback in massive MIMO FDD system. Multi-resolution paths and convolution factorization were first introduced to CSI feedback task and proved to be effective. Besides, an advanced training scheme was designed which successfully boosted the performance of the proposed CRNet. Experiments showed that the proposed CRNet greatly outperformed the state-of-the-art CsiNet with our advanced scheme under the same computational complexity.

\section*{Acknowledgment}
This work was supported in part by the National Key R\&D Program of China under Grant 2017YFE0112300 and Beijing National Research Center for Information Science and Technology under Grant BNR2019RC01014 and BNR2019TD01001.


\begin{thebibliography}{00}
\bibitem{01massivemimo} E. G. Larsson, O. Edfors, F. Tufvesson and T. L. Marzetta, ``Massive MIMO for next generation wireless systems,'' \textit{IEEE Communications Magazine}, vol. 52, no. 2, pp. 186-195, February 2014.
\bibitem{02massivemimo} L. Lu, G. Y. Li, A. L. Swindlehurst, A. Ashikhmin and R. Zhang, ``An Overview of Massive MIMO: Benefits and Challenges,'' \textit{IEEE Journal of Selected Topics in Signal Processing}, vol. 8, no. 5, pp. 742-758, Oct. 2014.
\bibitem{03massivemimo} M. S. Sim, J. Park, C. Chae and R. W. Heath, ``Compressed channel feedback for correlated massive MIMO systems,'' \textit{Journal of Communications and Networks}, vol. 18, no. 1, pp. 95-104, Feb. 2016.
\bibitem{05cs} P. Kuo, H. T. Kung and P. Ting, ``Compressive sensing based channel feedback protocols for spatially-correlated massive antenna arrays,'' 2012 \textit{IEEE Wireless Communications and Networking Conference (WCNC)}, Shanghai, 2012, pp. 492-497.
\bibitem{06resnet} K. He, X. Zhang, S. Ren and J. Sun, ``Deep Residual Learning for Image Recognition,'' 2016 \textit{IEEE Conference on Computer Vision and Pattern Recognition (CVPR)}, Las Vegas, NV, 2016, pp. 770-778.
\bibitem{07inceptionv1} C. Szegedy et al., ``Going deeper with convolutions,'' 2015 \textit{IEEE Conference on Computer Vision and Pattern Recognition (CVPR)}, Boston, MA, 2015, pp. 1-9.
\bibitem{09inceptionv3} C. Szegedy, V. Vanhoucke, S. Ioffe, J. Shlens and Z. Wojna, ``Rethinking the Inception Architecture for Computer Vision,'' 2016 \textit{IEEE Conference on Computer Vision and Pattern Recognition (CVPR)}, Las Vegas, NV, 2016, pp. 2818-2826.
\bibitem{10inceptionv4} Szegedy, Christian, et al. ``Inception-v4, inception-resnet and the impact of residual connections on learning,'' \textit{Thirty-First AAAI Conference on Artificial Intelligence}. 2017.
\bibitem{11reconnet} K. Kulkarni, S. Lohit, P. Turaga, R. Kerviche and A. Ashok, ``ReconNet: Non-Iterative Reconstruction of Images from Compressively Sensed Measurements,'' 2016 \textit{IEEE Conference on Computer Vision and Pattern Recognition (CVPR)}, Las Vegas, NV, 2016, pp. 449-458.
\bibitem{12dr2net} Yao, H., Dai, F., Zhang, S., Zhang, Y., Tian, Q., and Xu, C, ``Dr2-net: Deep residual reconstruction network for image compressive sensing,'' \textit{Neurocomputing}, 2019
\bibitem{13imagerecovery} C. Zou and F. Yang, ``Deep Learning Approach Based on Tensor-Train for Sparse Signal Recovery,'' \textit{IEEE Access}, vol. 7, pp. 34753-34761, 2019.
\bibitem{14csinet} C. Wen, W. Shih and S. Jin, ``Deep Learning for Massive MIMO CSI Feedback,'' \textit{IEEE Wireless Communications Letters}, vol. 7, no. 5, pp. 748-751, Oct. 2018.
\bibitem{15csinet_time_varying} T. Wang, C. Wen, S. Jin and G. Y. Li, ``Deep Learning-Based CSI Feedback Approach for Time-Varying Massive MIMO Channels,'' \textit{IEEE Wireless Communications Letters}, vol. 8, no. 2, pp. 416-419, April 2019.
\bibitem{16csinet_mapping} Y. Yang, F. Gao, G. Y. Li and M. Jian, ``Deep Learning based Downlink Channel Prediction for FDD Massive MIMO System,'' \textit{IEEE Communications Letters}, in press.
\bibitem{17csinet_mapping2} Z. Liu, L. Zhang and Z. Ding, ``Exploiting Bi-Directional Channel Reciprocity in Deep Learning for Low Rate Massive MIMO CSI Feedback,'' \textit{IEEE Wireless Communications Letters}, vol. 8, no. 3, pp. 889-892, June 2019.
\bibitem{18csinet_plus} J. Guo, C.-K. Wen, S. Jin, and G. Y. Li, ``Convolutional neural network based multiple-rate compressive sensing for massive MIMO CSI feedback: Design, simulation, and analysis,'' \textit{arXiv preprint arXiv:1906.06007}, 2019
\bibitem{20sgdr} Loshchilov, Ilya, and Frank Hutter, ``SGDR: STOCHASTIC GRADIENT DESCENT WITH RESTARTS,'' \textit{Learning} 10: 3.
\bibitem{19Cost2100} L. Liu et al., ``The COST 2100 MIMO channel model,'' \textit{IEEE Wireless Communications}, vol. 19, no. 6, pp. 92-99, December 2012.
\end{thebibliography}
\end{document}